\documentclass[twocolumn,letterpaper,10pt]{article}
\usepackage[letterpaper,margin=0.75in]{geometry} 
\usepackage[font=footnotesize]{caption}

\usepackage{titlesec}
\titleformat{\section}{\normalfont\fontsize{10}{12}\bfseries\uppercase}{\thesection}{1em}{}
\titleformat{\subsection}{\normalfont\fontsize{10}{12}\bfseries}{\thesubsection}{1em}{}
\titleformat{\subsubsection}{\normalfont\fontsize{10}{12}\itshape}{\thesubsubsection}{1em}{}

\renewenvironment{abstract}{
  \small\noindent\textbf{Abstract—}\ignorespaces
}{\par\bigskip}

\usepackage{etoolbox}
\patchcmd{\thebibliography}{\section*{\refname}}{\section*{\refname}\footnotesize}{}{}

\usepackage[utf8]{inputenc}
\usepackage{amsmath, amssymb}
\usepackage{graphicx}
\usepackage{authblk}
\usepackage{booktabs}
\usepackage{threeparttable}
\usepackage{lipsum}

\usepackage{pifont}
\newcommand{\cmark}{\ding{51}} 
\newcommand{\xmark}{\ding{55}} 
\usepackage{url}

\title{DHAuDS: A Dynamic and Heterogeneous Audio Benchmark for Test-Time Adaptation}
\author{Weichuang~Shao}
\author{Iman~Yi Liao}
\author{Tomas~Henrique~Bode~Maul}
\author{Tissa~Chandesa}
\affil{School of Computer and Mathematical Sciences, University of Nottingham Malaysia, Semenyih, Malaysia \\ 
\texttt{andyshao90@gmail.com,\{Iman.Liao, Tomas.Maul, Tissa.Chandesa\}@nottingham.edu.my}}
\date{}

\begin{document}

\maketitle

\begin{abstract}
    Existing Test-time Adaptation (TTA) studies rely heavily on static and homogeneous corruption protocols, such as ImageNet-C and CIFAR-10-C/100-C, leading to inconsistent evaluation settings and potentially inflated robustness estimates that are compared with real-world situations. TTA lacks a standardized evaluation infrastructure capable of modeling realistic heterogeneous acoustic degradation. We introduce DHAuDS, a standardized benchmark suite for evaluating audio classification TTA robustness under dynamic corruption severity and heterogeneous noise mixtures. Rather than proposing a new TTA algorithm, DHAuDS focuses on exposing robustness limitations that remain hidden under conventional fixed-noise evaluation protocols. 
\end{abstract}

\vspace{1em}
\noindent\textbf{Keywords:} Deep Learning, Audio Classification, Audio Benchmark, Test-time Adaptation
\vspace{1em}

\section{Introduction}
While several studies have proposed Test-time Adaptation (TTA) techniques for audio classification and Automatic Speech Recognition (ASR), a consistent evaluation protocol across domains remains unavailable~\cite {lin2022listen,kim2023sgem,lin2024continual,Amiri2024PathologySpeechDetection,shao2025investigation}. Consequently, existing works employ distinct data and noise configurations, which complicates the direct comparison of their results~\cite {lin2022listen,kim2023sgem,lin2024continual,Amiri2024PathologySpeechDetection,shao2025investigation}. Further, the research presented in Table~\ref{tab:alg_sup} (excluding TTAAPSD) uses the identical corrupted test set for both adaptation and evaluation, which makes it difficult to detect model overfitting. In contrast, DHAuDS introduces divergence between the adaptation and evaluation phases by independently corrupting the test sets, thereby emphasizing robustness analysis rather than mere correctness. 

The goal of DHAuDS is not to introduce another adaptation algorithm, but to establish a reproducible and standardized robustness evaluation protocol for audio-TTA research to determine whether we can overcome a critical limitation in existing TTA evaluation protocols: reliance on static corruption severity and homogeneous noise conditions. Similar to the role of ImageNet-C~\cite{hendrycks2019benchmarking} in visual robustness research, DHAuDS aims to support fair cross-method comparison under realistic acoustic domain shifts.
Our contributions are:
\begin{enumerate}
    \item We reveal a fundamental limitation of existing audio-TTA evaluation protocols, namely their reliance on static corruption severity and homogeneous noise conditions, which may overestimate robustness.
    \item We introduce DHAuDS, a standardized benchmark suite spanning speech, vocal, urban, and bioacoustic classification tasks.
    \item We propose dynamic corruption severity, heterogeneous noise mixtures, and independent adaptation/evaluation sets to better approximate realistic acoustic domain shifts.
\end{enumerate}

\begin{table*}[t]
    \centering
    \caption{Support for dynamic corruption levels (DyN) and heterogeneous noise (Heter) in existing TTA studies}
    \label{tab:alg_sup}
    \footnotesize
    \begin{tabular}{l|cccl}
        \toprule
        Algorithm & Application Domain & DyN & Heter & Notes / Dataset Example \\
        \midrule
        SUTA~\cite{lin2022listen} & Audio (ASR) & \xmark & \xmark & Gaussian noise or CHiME-3~\cite{barker2017third} noise on Librispeech~\cite{panayotov2015librispeech} \\

        SGEM~\cite{kim2023sgem} & Audio (ASR) & \xmark & \xmark & Fixed 10 dB SNR, MS-SNSD~\cite{reddy2019scalable} \\

        DSUTA~\cite{lin2024continual} & Audio (ASR) & \xmark & \xmark & Fixed 5 dB SNR, MS-SNSD \\

        TTAAPSD~\cite{Amiri2024PathologySpeechDetection} & Audio (Pathological Speech) & \xmark & \cmark & Multiple noises from QUT-NOISE~\cite{dean2010qut} and DEMAND~\cite{thiemann2013demand} \\
        & & & & but fixed SNR per test\\

        CoNMix++~\cite{shao2025investigation} & Audio (Classification) & \xmark & \xmark & Three noises, fixed SNR (3 or 10 dB)  \\

        L‑TTA~\cite{shin2024tta} & Image (Classification) & \xmark & \xmark & ImageNet-C~\cite{hendrycks2019benchmarking}, CIFAR-10-C~\cite{hendrycks2019benchmarking}, fixed severity level \\

        IST~\cite{ma2024improved} & Image (Classification) & \xmark & \xmark & ImageNet-C~\cite{hendrycks2019benchmarking}, CIFAR-10-C~\cite{hendrycks2019benchmarking}, fixed severity level \\

        ResNeXt~\cite{hendrycks2021many} & Image (Classification) & \xmark & \cmark & ImageNet-R~\cite{hendrycks2021many}; Natural domain shift, e.g, Paintings and Cartoons \\
        
        AlexNet~\cite{wang2019learning} & Image (Classification) & \xmark & \xmark & ImageNet-Sketch~\cite{wang2019learning}; Pencoil drawing that discard color and texture. \\
        DHAuDS & Audio (Classification) & \cmark & \cmark & Dynamic severity level and heterogeneous noise type per experiment\\
        \bottomrule
    \end{tabular}
\end{table*}

\begin{table*}[t]
    \centering
    \caption{DHAuDS corruption settings}
    \label{tab:noise_type_setting}
    \footnotesize
    \begin{threeparttable}
    \begin{tabular}{l|cc|l}
        \toprule
        Category-level & \multicolumn{2}{c|}{Dynamic SNR Score} & Heterogeneous Noise-type \\
        \midrule
        &Severity Range & Step & \\ 
        \midrule
        WHN-L1 & [6 dB, 7 dB] & 0.5 dB & Gaussian, Random\\
        WHN-L2 & [5 dB, 7 dB] & 0.5 dB & Gaussian, Random\\
        ENQ-L1 & [5 dB, 6 dB] & 0.5 dB & HOME, REVERB, STREET\\
        ENQ-L2 & [5 dB, 7 dB] & 0.5 dB & HOME, REVERB, STREET, CAFE, CAR\\
        END1-L1 & [5 dB, 6 dB] & 0.5 dB & NFIELD, PRESTO, TCAR, OOFFICE \\
        END1-L2 & [5 dB, 7 dB] & 0.5 dB & NFIELD, PRESTO, TCAR, OOFFICE, STRAFFIC, DKITCHEN\\
        END2-L1 & [5 dB, 6 dB] & 0.5 dB & DLIVING, OHALLWAY, SPSQUARE, TMETRO\\
        END2-L2 & [5 dB, 7 dB] & 0.5 dB & DLIVING, OHALLWAY, SPSQUARE, TMETRO, NRIVER, PSTATION\\
        ENSC-L1 & [5 dB, 6 dB] & 0.5 dB & exercise\_bike, running\_tap, white\_noise, pink\_noise\\
        ENSC-L2 & [5 dB, 7 dB] & 0.5 dB & exercise\_bike, running\_tap, white\_noise, pink\_noise, doing\_the\_dishes, dude\_miaowing\\
        TST-L1 & [-6\%, -4\%] $\cup$ [4\%, 6\%] & 1\% & N/A\\
        TST-L2 & [-12\%, -8\%] $\cup$ [8\%, 12\%] & 1\% & N/A\\
        PSH-L1 & [-5 st, -4 st] $\cup$ [4 st, 5 st] & 1 st & N/A\\
        PSH-L2 & [-7 st, -5 st] $\cup$ [5 st, 7 st] & 1 st & N/A\\
        \bottomrule
    \end{tabular}
    \scriptsize
    \begin{tablenotes}
        \item[1] The unit of PSH is the pitch shift amount - semitone (st).
        \item[2] We select the Signal-to-Noise Ratio (SNR) score and cut the audio clips randomly.
    \end{tablenotes}
    \end{threeparttable}
\end{table*}
\section{Related Work}
The current evaluation protocols of TTA, which self-corrects at inference time using unlabeled test sets, in audio are fragmented. Unlike Computer Vision benchmarks (e.g., ImageNet-C~\cite{hendrycks2019benchmarking} and CIFAR-10-C/100-C~\cite{hendrycks2019benchmarking}), which offer standardized severities, audio TTA research lacks a unified protocol. In Table~\ref{tab:alg_sup}, SUTA~\cite{lin2022listen}, SGEM~\cite{kim2023sgem}, DSUTA~\cite{lin2024continual}, TTAAPSD~\cite{Amiri2024PathologySpeechDetection}, and CoNMix++~\cite{shao2025investigation} leverage different evaluation protocols. Additionally, they rely on fixed Signal-to-Noise Ratios (SNRs) or on a single noise type per experiment, failing to capture the dynamic (varying severity) and heterogeneous (mixed sources) nature of real-world acoustic degradation. 

In contrast, computer-vision research has benefited from standardized robustness benchmarks such as ImageNet-C and CIFAR-10-C/-100-C, which evaluate models under defined corruption levels~\cite{hendrycks2019benchmarking,croce2021robustbench}. However, these visual perturbations, such as contrast or brightness shifts, \textbf{do not translate naturally} into the acoustic domain. Furthermore, they generally use one fixed noise intensity and a singular noise type per test (see Table~\ref{tab:alg_sup}), \textbf{lacking the dynamic and composite conditions} typical of real-world recordings~\cite{shin2024tta,ma2024improved}.

These factors motivate the development of the DHAuDS benchmark, which models audio degradations that vary both in type and intensity, offering a more faithful simulation of domain shift during inference.

\section{Methodology}
The DHAuDS benchmark aims to replicate the variability of real-world acoustic conditions, creating challenging yet controlled settings for assessing TTA methods. In contrast to previous works~\cite {lin2022listen,kim2023sgem,lin2024continual,Amiri2024PathologySpeechDetection,shao2025investigation,hendrycks2019benchmarking} that rely on a single fixed noise level, DHAuDS applies \textbf{dynamic SNR score selection} and \textbf{heterogeneous noise mixtures} to reflect natural acoustic diversity.

\subsection{Design Principles}
DHAuDS is built upon three principles: dynamic corruption severity, heterogeneous noise conditions, and independent adaptation/evaluation corruption instances. Together, these principles aim to better reflect real-world acoustic variability and provide a more rigorous robustness evaluation protocol for audio TTA.

\subsection{Benchmark Overview}
DHAuDS consists of four benchmarks: UrbanSound8K-C (US8-C), SpeechCommands V2-C (SC2-C), VocalSound-C (VS-C), and ReefSet-C (RS-C) (see Table~\ref{tab:dhauds_benchmark}). Each benchmark is divided into two sets: the adaptation set and the evaluation set. Both sets contain seven corruption categories and two levels of severity.
\begin{table}[ht]
    \centering
    \caption{DHAuDS benchmark (BM) information}
    \label{tab:dhauds_benchmark}
    \footnotesize
    \begin{tabular}{l|ccc}
        \toprule
        BM & Total Size & Per Set & Per Cat.-lvl. \\
        \midrule
        SC2-C & 308,140 & 154,070 & 11,005 \\
        VS-C & 100,548 & 50,274 & 3,591 \\
        US8-C & 19,672 & 9,836 & 2,459 \\
        RS-C & 479,836 & 239,918 & 17,137 \\
        \bottomrule
    \end{tabular}
\end{table}
As shwon in Table~\ref{tab:noise_type_setting}, corruption categories are defined: \emph{(1)} White Noise (WHN); \emph{(2)-(5)} Environmental noise drawn from QUT-NOISE~\cite{dean2010qut} (ENQ), DEMAND~\cite{thiemann2013demand} (END1 and END2), and SpeechCommands V2~\cite{warden2018speech} (ENSC); \emph{(6)} Time Stretching (TST) via random tempo adjustments; and \emph{(7)} Pitch Shifting (PSH). Two levels are defined: L1 (standard) and L2 (challenging), where L2 applies broader corruption ranges and more complex noise combinations. \textit{Notably, each sample is affected by only one type of noise, though the noise types differ between samples in the same corruption category}.

\begin{table}[ht]
    \caption{Audio set information}
    \label{tab:dataset}
    \centering
    \footnotesize
    \begin{tabular}{l|ccccl}
        \toprule
        Set & Type & Len. & Bal. & Rate & Size \\
        \midrule
        SC2 & speech & 1 s & Yes & 16 kHz & 105829\\
        VS & vocal & 12 s & Yes & 16 kHz & 20977\\
        US8 & urban & 4 s & No & 44.1 kHz & \textbf{8732}\\
        RS & bioacoustic & 1.88 s & No & 16 kHz & 57074\\
        \bottomrule
    \end{tabular}
\end{table}
We selected four audio classification sets, UrbanSound8K (US8)~\cite{salamon2014dataset}, SpeechCommands V2 (SC2)~\cite{warden2018speech}, VocalSound (VS)~\cite{gong_vocalsound}, and ReefSet (RS)~\cite{williams2025using}, to represent different audio types, lengths, sample rates, and whether they are balanced, allowing us to evaluate TTA performance across various conditions. As shown in Table~\ref{tab:dataset}, four audio sets are used to generate our benchmarks: US8-C, SC2-C, VS-C, and RS-C. \textit{Notably, US8-C excludes ENQ, END1, and END2 because their class types overlap with noise types in QUT-NOISE and DEMAND, such as street, car, traffic, and station noise}.

To ensure \textit{robustness}, we utilize adaptation sets for model adaptation and evaluation sets for analyzing performance. To create unrelated adaptation and evaluation sets, we employ different random seeds: \textbf{seed 2025} for the adaptation set and \textbf{seed 123456} for the evaluation set. 
To ensure \textit{reproducibility}, both adaptation and evaluation sets are publicly released for future research.

In the experiments, we report the F1-score for US8 (F1-score better than accuracy on the imbalance set), ROC-AUC for RS (RS publisher choice), and Top-1 Accuracy for the others.

\subsection{Corruption Protocol}
\subsubsection{Dynamic SNR score}
Crucially, we implement a dynamic severity mechanism. Rather than a fixed SNR, corruption intensity is sampled randomly from defined ranges (see Table~\ref{tab:noise_type_setting}). This prevents models from overfitting to a specific noise floor.
\subsubsection{Heterogenous noise mixtures}
The noise corruptions are constructed from multiple publicly available datasets to ensure diversity and realism. Specifically, \textbf{ENQ} utilizes the complete set of 20 audio recordings (approximately 818 minutes in total), encompassing various ambient environments such as CAFE, CAR, HOME, REVERB, and STREET. Environmental noise is divided into two subsets, \textbf{END1} and \textbf{END2}, each containing 96 recordings (16 per noise type) with a total duration of roughly 480 minutes, capturing a broad range of indoor and outdoor acoustic scenes. \textbf{ENSC} employs all six short background noise clips (approximately 399 seconds in total) from SC2, including sounds such as doing\_the\_dishes, running\_tap, pink\_noise, white\_noise, exercise\_bike, and dude\_miaowing. For \textbf{TST}, an exception is made for short 1-second datasets like SC2, where the slowing-down operation is omitted to prevent truncation that may remove critical speech content. \textbf{PSH} adjusts the pitch upward or downward by several semitone steps without changing the temporal duration of the signal~\cite{morrison2021neural,wu2021quasi}.

\section{Experiment}
\subsection{Models for Evaluation}
We aim to facilitate an objective evaluation across DHAuDS benchmarks. We adopt three representative architectures, AMAuT~\cite{shao2025amaut}, HuBERT~\cite{hsu2021hubert}, and PANNs~\cite{kong2020panns}, to evaluate DHAuDS benchmarks (as shown in Table~\ref{tab:model_inf}).
\begin{table}[ht]
    \centering
    \caption{Model Information}
    \label{tab:model_inf}
    \small
    \begin{tabular*}{\linewidth}{@{\extracolsep\fill}lcl}
        \toprule
        Modle & Architecture & Input Type \\
        \midrule
        AMAuT & 1D CNN-Transformer & Mel-spectrogram \\ 
        HuBERT & 1D CNN-Transformer & Raw audio \\
        PANNs & 1D \& 2D CNNs & Wavgram log-Mel \\
        \bottomrule
    \end{tabular*}
\end{table}
The DHAuDS benchmarks present a challenge due to inconsistent audio lengths—specifically 1 s, 1.88 s, 4 s, and 12 s—and varying sample rates of 16 kHz and 44.1 kHz (see Table~\ref{tab:dataset}). \textit{All three models can process dynamic audio lengths and successfully execute all experiments in the DHAuDS benchmarks}. 

\subsection{Test-Time Adaptation (TTA) Strategy}\label{ssec:tta}
Existing TTA methods have primarily been developed for image classification, ASR, or task-specific speech applications rather than general audio classification~\cite{shin2024tta,ma2024improved,wang2019learning,lin2022listen,kim2023sgem,lin2024continual}. Their adaptation objectives and architectural assumptions are often tightly coupled to their original domains, making direct deployment on the diverse audio-classification benchmarks in DHAuDS difficult without substantial redesign and revalidation. Therefore, we employ TENT, TTN~\cite{wang2021tent}, and a representative CoNMix-based adaptation objective solely to demonstrate benchmark behavior rather than to establish a new state-of-the-art method. In particular, TENT and TTN require a BatchNorm component that cannot be reused in HuBERT. 
\begin{table*}[t]
    \centering
    \caption{Adaptation performance comparison on evaluation set of SC2-C, VS-C, RS-C, and US8-C benchmarks (BMs)}
    \label{tab:tta_perf}
    \begin{threeparttable}
    \footnotesize
    \begin{tabular}{lcc|cccccc}
        \toprule
        Model & BM & Level & No Adaptation & CoNMix-based TTA & TENT & TTN \\ 
        \midrule
        AMAuT & SC2-C & L1 & $.8289\pm.0000$ & $.8888\pm.0018$ & $.3180\pm.0001\downarrow$ & $.8630\pm.0006$ \\
        HuBERT & SC2-C & L1 & $.9259\pm.0000$ & $.9513\pm.0006$ & N/A & N/A \\
        PANNs & SC2-C & L1 & $.7478\pm.0000$ & $.8090\pm.0007$ & $.0388\pm.0001\downarrow$ & $.2361\pm.0115\downarrow$\\
        AMAuT & SC2-C & L2 & $.8049\pm.0000$ & $.8713\pm.0023$ & $.3108\pm.0001\downarrow$ & $.8371\pm.0010$ \\
        HuBERT & SC2-C & L2 & $.9084\pm.0000$ & $.9486\pm.0007$ & N/A & N/A \\
        PANNs & SC2-C & L2 & $.7273\pm.0001$ & $.7938\pm.0011$ & $.0390\pm.0002\downarrow$ & $.2347\pm.0136\downarrow$\\
        \midrule
        AMAuT & VS-C & L1 & $.7838\pm.0000$ & $.8734\pm.0028$ & $.8409\pm.0015$ & $.8582\pm.0044$ \\
        HuBERT & VS-C & L1 & $.7419\pm.0001$ & $.9083\pm.0022$ & N/A & N/A \\
        PANNs & VS-C & L1 & $.7972\pm.0000$ & $.8945\pm.0011$ & $.7824\pm.0054\downarrow$ & $.5917\pm.0066\downarrow$ \\
        AMAuT & VS-C & L2 & $.7767\pm.0001$ & $.8668\pm.0027$ & $.8356\pm.0017$ & $.8506\pm.0025$ \\
        HuBERT & VS-C & L2 & $.7401\pm.0001$ & $.9051\pm.0021$ & N/A & N/A \\
        PANNs & VS-C & L2 & $.7932\pm.0000$ & $.8904\pm.0010$ & $.7741\pm.0049\downarrow$ & $.5880\pm.0047\downarrow$ \\
        \midrule
        AMAuT & RS-C & L1 & $.7429\pm.0000$ & $.8791\pm.0020$ & $.4890\pm.0016\downarrow$ & $.8781\pm.0011$ \\
        HuBERT & RS-C & L1 & $.7559\pm.0000$ & $.8703\pm.0010$ & N/A & N/A \\
        PANNs & RS-C & L1 & $.8130\pm.0000$ & $.9269\pm.0009$ & $.5229\pm.0001\downarrow$ & $.9096\pm.0031$ \\
        AMAuT & RS-C & L2 & $.7359\pm.0000$ & $.8803\pm.0017$ & $.4872\pm.0020\downarrow$ & $.8686\pm.0012$ \\
        HuBERT & RS-C & L2 & $.7540\pm.0000$ & $.8619\pm.0016$ & N/A & N/A \\
        PANNs & RS-C & L2 & $.8069\pm.0000$ & $.9209\pm.0013$ & $.5234\pm.0001\downarrow$ & $.9024\pm.0025$ \\
        \midrule
        AMAuT & US8-C & L1 & $.5654\pm.0001$ & $.6882\pm.0072$ & $.4107\pm.0003\downarrow$ & $.6568\pm.0052$ \\
        HuBERT & US8-C & L1 & $.5539\pm.0000$ & $.6758\pm.0128$ & N/A & N/A \\
        PANNs & US8-C & L1 & $.7122\pm.0000$ & $.8412\pm.0025$ & $.5406\pm.0030\downarrow$ & $.8134\pm.0053$ \\
        AMAuT & US8-C & L2 & $.5594\pm.0000$ & $.6760\pm.0070$ & $.4004\pm.0002\downarrow$ & $.6413\pm.0039$ \\
        HuBERT & US8-C & L2 & $.5458\pm.0000$ & $.6649\pm.0052$ & N/A & N/A \\
        PANNs & US8-C & L2 & $.7078\pm.0000$ & $.8321\pm.0028$ & $.5325\pm.0025\downarrow$ & $.8076\pm.0040$ \\
        \bottomrule
    \end{tabular}
    \scriptsize
    \begin{tablenotes}
        \item[1] To evaluate robustness, all experiments were repeated across five random seeds. We report the mean and standard deviation aggregated across corruption categories.  
    \end{tablenotes}
    \end{threeparttable}
\end{table*}

As for CoNMix-based TTA, we adopt Nuclear-Norm Maximization ($\mathcal{L}_{ncm}$) from CoNMix~\cite{kumar2023conmix} as the primary adaptation objective. Because the remaining CoNMix components were developed for fixed-resolution image-classification settings, we replace them with an audio-specific consistency loss that encourages stable predictions across temporally shifted views of the same sample. The model is then adjusted using $\mathcal{L} = \mathcal{L}_{ncm} + \lambda \mathcal{L}_{con}$ for each corruption category listed in Table \ref{tab:noise_type_setting}, treating them independently.

Equation (1) defines the consistency loss applied to predictions from two augmented views generated by left-shifted ($x_l$) or right-shifted ($x_r$) views of each test sample.
\begin{equation}\label{eq:cst_ls}
    \mathcal{L}_{con} = \frac{1}{B} \sum_{i=1}^B \sum_{c=1}^C \big|\big|\hat{p}_{i,c}(x_l) - \hat{p}_{i,c}(x_r)\big|\big|_2^2
\end{equation}
where $B$ is the batch size, $C$ is the number of classes, and $\hat{p}_{i,c}(x)$ is the predicted probability of class $c$ for sample $i$.

\subsection{Cross-Domain and Model Robustness Under DHAuDS}
To ensure the robustness of performance evaluation in Table~\ref{tab:tta_perf}, all experiments were repeated using \textit{five random seeds} (2025, 123456, 654321, 891011, and 111098). All experiments use a batch size of 32 for evaluation, excluding adaptation, as \textit{different sizes may affect prediction performance}. Additionally, all experiments adapted on the adaptation set while evaluating on the evaluation sets.

As shown in Table~\ref{tab:tta_perf}, both TENT and TTN demonstrate negative optimization effects on the DHAuDS benchmarks. Specifically, TENT triggers negative optimization across all DHAuDS benchmarks, while TTN only exhibits this issue between SC2-C and VS-C. CoNMix-based TTA demonstrates improved performance in all experiments compared to the others in Table~\ref{tab:tta_perf}. The presence of negative optimization highlights the robustness gaps that previous evaluation protocols fail to address, as negative optimization is not prominently observed under the evaluation protocols reported in the original studies~\cite{wang2021tent}.

Combining the discussion from this section with subsection~\ref{ssec:tta}, the previous TTA methods face challenges related to negative optimization effects or to designing an audio classification TTA method across the DHAuDS benchmarks. However, \textit{one of the motivations for publishing DHAuDS is to support future evaluations of developed TTA methods for audio classification}.

\section{Ablation Study}
\subsection{Impact of Dynamic and Heterogeneous Protocols on Robustness Evaluation}
To illustrate the implications of the dynamic and heterogeneous corruption strategy, we compare performance using static or dynamic SNR score, with noise that is either homogeneous or heterogeneous.
\begin{table}[t]
    \centering
    \caption{A comparison of prediction accuracy before and after adaptation}
    \label{tab:fix_dynamic_comp}
    \begin{threeparttable}
    \scriptsize
    \begin{tabular}{lc|cc}
        \toprule
        Model & Group & Performance & $\Delta$ \\
        \toprule
        \multicolumn{4}{l}{SC2/SC2-C} \\
        \midrule
        AMAuT & S\&HO & $.8883\pm.0000\rightarrow.9107\pm.0013$ & \textbf{.0224$\pm$.0000} \\
        AMAuT & D\&HE & $.9021\pm.0000\rightarrow.9168\pm.0026$ & $.0148\pm.0026$\\
        HuBERT & S\&HO & $.9554\pm.0000\rightarrow.9635\pm.0002$ & \textbf{.0081$\pm$.0000}\\
        HuBERT & D\&HE & $.9587\pm.0000\rightarrow.9640\pm.0004$ & $.0054\pm.0004$\\
        PANNs & S\&HO & $.8124\pm.0000\rightarrow.8551\pm.0005$ & \textbf{.0427$\pm$.0005}\\
        PANNs & D\&HE & $.8353\pm.0001\rightarrow.8476\pm.0004$ & $.0123\pm.0003$\\
        \toprule
        \multicolumn{4}{l}{VS/VS-C} \\
        \midrule
        AMAuT & S\&HO & $.7683\pm.0000\rightarrow.8804\pm.0033$ & \textbf{.1121$\pm$.0000}\\
        AMAuT & D\&HE & $.8063\pm.0002\rightarrow.8850\pm.0024$ & $.0788\pm.0023$\\
        HuBERT & S\&HO & $.7070\pm.0000\rightarrow.9167\pm.0027$ & \textbf{.2097$\pm$.0000}\\
        HuBERT & D\&HE & $.7877\pm.0002\rightarrow.9136\pm.0016$ & $.1259\pm.0016$\\
        PANNs & S\&HO & $.8708\pm.0000\rightarrow.9021\pm.0018$ & \textbf{.0313$\pm$.0018}\\
        PANNs & D\&HE & $.8805\pm.0000\rightarrow.9032\pm.0012$ & $.0227\pm.0012$\\
        \toprule
        \multicolumn{4}{l}{RS/RS-C} \\
        \midrule
        AMAuT & S\&HO & $.6558\pm.0000\rightarrow.8844\pm.0021$ & \textbf{.2286$\pm$.0000}\\
        AMAuT & D\&HE & $.7491\pm.0000\rightarrow.8782\pm.0024$ & $.1291\pm.0024$\\
        HuBERT & S\&HO & $.7100\pm.0000\rightarrow.8080\pm.0037$ & \textbf{.0980$\pm$.0000}\\
        HuBERT & D\&HE & $.7833\pm.0000\rightarrow.8526\pm.0025$ & $.0693\pm.0025$\\
        PANNs & S\&HO & $.7728\pm.0000\rightarrow.9025\pm.0027$ & \textbf{.1297$\pm$.0027}\\
        PANNs & D\&HE & $.8361\pm.0000\rightarrow.9074\pm.0015$ & $.0714\pm.0015$\\
        \bottomrule
    \end{tabular}
    \scriptsize
    \begin{tablenotes}
        \item[1] To evaluate robustness, all experiments were repeated across five random seeds. 
        \item[2] Before adaptation$\rightarrow$After adaptation 
        \item[3] We do not analyze US8/US8-C since it exludes END2-L2. 
    \end{tablenotes}
    \scriptsize
    \end{threeparttable}
\end{table}
In Table~\ref{tab:fix_dynamic_comp}, S\&HO denotes  5 dB SNR + PSTATION noise on the test set of SC2, VS, and RS. D\&HE presents END2-L2 on SC2-C, VS-C, and RS-C. Both of the groups, S\&HO and D\&HE, adopt the CoNMix-based TTA. 

As shown in Table~\ref{tab:fix_dynamic_comp}, the S\&HO group presents a larger improvement than the D\&HE group after adaptation, which indicates less complexity and fewer challenges. Furthermore, compared with the S\&HO group, the D\&HE group setup introduces greater variability and diversity in acoustic corruption, resulting in a more challenging robustness evaluation scenario.

In conclusion, static-SNR homogeneous evaluation produces optimistic conclusions, dynamic-SNR heterogeneous corruption reveals reduced generalization; therefore, static and homogeneous protocols are insufficient.


\subsection{Complexity of Dynamic and Heterogeneous Protocols}
Some of DHAuDS corruption categories and levels pose significant challenges, such as END1-L1/-L2 and END2-L1/-L2. 
\begin{table}[ht]
    \centering
    \caption{Sample size vs Non-repeating noise clips in END1-L2}
    \label{tab:sample_size_vs_clip}
    \footnotesize
    \begin{threeparttable}
    \begin{tabular}{l|cccl}
        \toprule
        BM & Size & Length & Non-repeating Clips & Size/Clips \\
        \midrule
        SC2-C & 11,005 & 1 s & $\frac{480 \times 60}{1} \times 5 = 144,000$ & 7.6\% \\
        VS-C & 3,591 & 12 s & $\frac{480 \times 60}{12} \times 5 = 12,000$ & \textbf{29\%} \\
        RS-C & 17,137 & 1.88 s & $\frac{480 \times 60}{1.88} \times 5 \approx 76,595$ & 22.4\% \\
        \bottomrule
    \end{tabular}
    \scriptsize
    \begin{tablenotes}
        \item US8-C does not include END1-L2, so we won't make a comparison here. The sample size of the test set on US8 is 2,459, and the evaluation result is consistent with the others.
    \end{tablenotes}
    \end{threeparttable}
\end{table}
For example, END1-L2 comprises 5 SNR scores and 480 minutes of background noise across various noise types. In Table~\ref{tab:sample_size_vs_clip}, END1-L2 used only 29\% of the total set of non-repeating clips across all benchmarks of DHAuDS, which implies that every sample is corrupted differently.

\subsection{Impact of Momentum in TTA}
\begin{figure}[ht]
    \centering
    \includegraphics[width=0.7\linewidth]{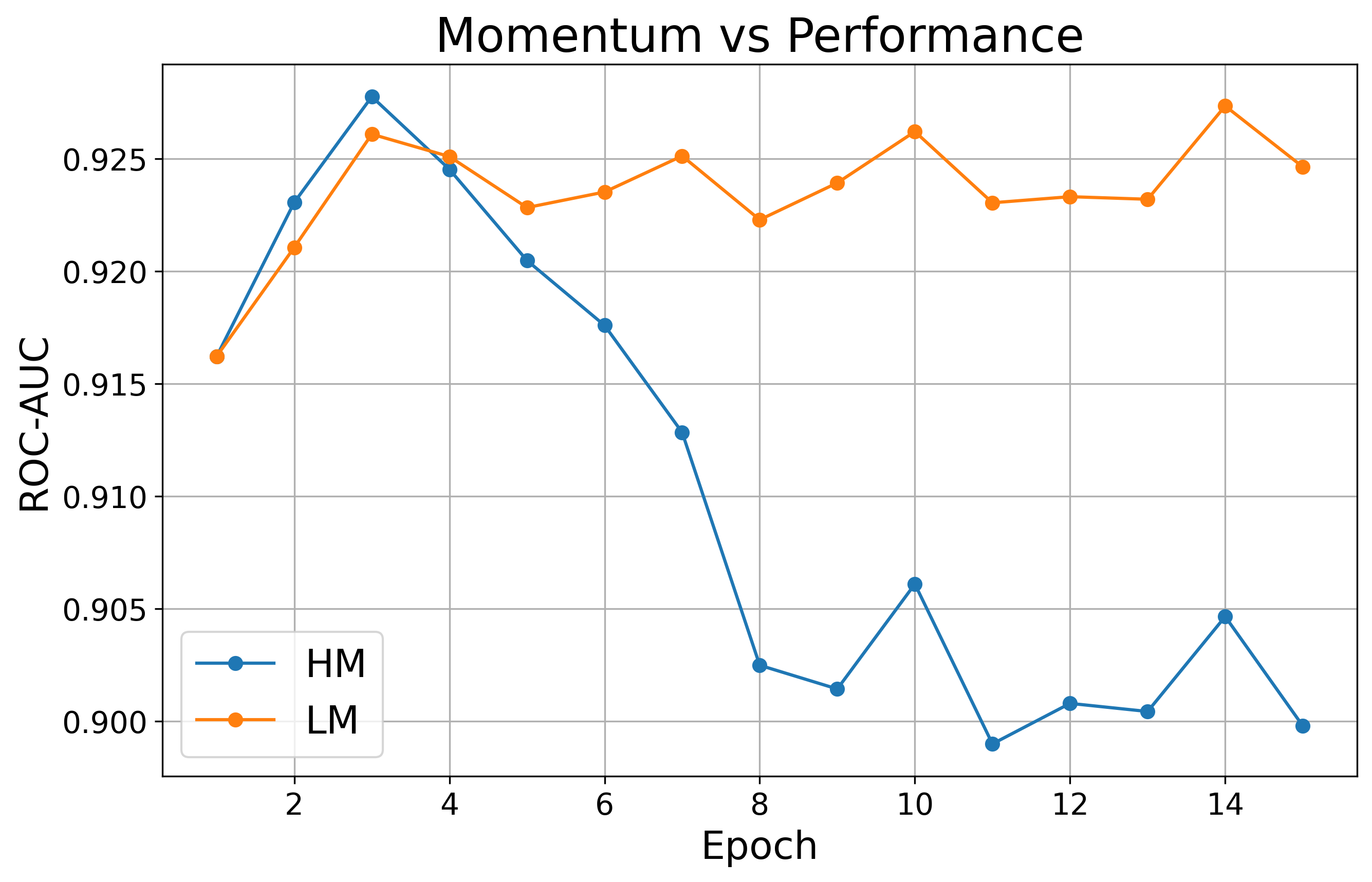}
    \caption{Comparison of ROC–AUC performance between high-momentum (HM = 0.90) and low-momentum (LM = 0.70) settings when performing CoNMix-based TTA on AMAuT under WHN-L1 of RS-C. All other hyperparameters remain identical.}
    \label{fig:IoM_AuT-RS_WHN-L1}
\end{figure}
Our experiments indicate that adopting a lower momentum value ($0.70-0.75$) stabilizes TTA, reducing the decline in performance that often follows early accuracy gains. As shown in Figure~\ref{fig:IoM_AuT-RS_WHN-L1}, a lower momentum (0.70) maintains prediction stability compared to a higher value (0.90) when applying CoNMix-based TTA on AMAuT under WHN-L1 of RS-C.


\section{Conclusion}
Existing TTA evaluations (see Table~\ref{tab:alg_sup}) rely heavily on fixed severity or homogeneous corruptions, or both, whereas DHAuDS introduces dynamic severity, heterogeneous corruptions, and adaptation/evaluation separation to expose robustness limitations that may otherwise remain hidden.
The DHAuDS evaluation spans diverse audio domains, recording characteristics, corruption conditions, and model architectures to provide a systematic and rigorous assessment of robustness. We expect DHAuDS to serve as a standardized robustness evaluation protocol that facilitates the development and fair comparison of future audio classification TTA methods.

\section*{Acknowledgments}
AI-assisted tools were used solely to improve the quality and readability of language, rather than to generate research ideas, experimental results, or analyses. The complete codebase and benchmark datasets are publicly available at: \url{https://github.com/Andy-Shao/DHAuDS}

\bibliographystyle{ieeetr}
\bibliography{reference}

\appendix

\section{Experiment Settings}
\subsection{Modifications in PANNs}
The benchmarks of DHAuDS specifically target the test sets of four audio datasets: SC2~\cite{warden2018speech}, RS~\cite{williams2025using}, VS~\cite{gong_vocalsound}, and US8~\cite{salamon2014dataset}. DHAuDS is necessary to process the training mode during the training phase of these four datasets before adapting to the four benchmarks. However, the training mode of PANNs is designed to adapt to AudioSet~\cite{gemmeke2017audio}, which contains approximately 2 million samples in its training set. To enhance robustness and mitigate overfitting, PANNs incorporate Dropout layers in each CNN block. In contrast, the training sets for SC2, RS, VS, and US8 consist of 84,843, 39,937, 15,531, and 6,273 samples, respectively, which is significantly fewer than the 2 million samples in AudioSet. Dropout layers in PANNs tend to diminish prediction performance because the sample size of the training set is insufficient. Therefore, we have decided to eliminate the Dropout layers from each CNN block, retaining only the final Dropout layer.

\subsection{AMAuT Pre-training for US8}
The US8~\cite{salamon2014dataset} dataset comprises 10 classes of urban audio content, with audio segments under 4 seconds and varying sample rates. All recordings are resampled to 44.1 kHz. Because AMAuT has over 99 million parameters, it requires a large training set ($\ge$15,000 samples)~\cite{shao2025amaut}. Since US8 contains only 8,732 samples (see Table~\ref{tab:dataset}), it is insufficient for training AMAuT from scratch.

To resolve this, pre-training is performed using CochlScene~\cite{jeong2022cochlscene}, which meets the requirements of: \emph{(1)} Sample size = 76,115; \emph{(2)} Sample rate = 44.1 kHz; and \emph{(3)} Urban audio content (e.g., \textit{bus}, \textit{car}, \textit{subway station}, \textit{café}). AMAuT is first trained on CochlScene, and the \textbf{pre-trained parameters} are then \textbf{transferred to US8} for fine-tuning.


\end{document}